\documentclass[sigplan]{acmart}
\acmSubmissionID{97}
\renewcommand\footnotetextcopyrightpermission[1]{} 
\AtBeginDocument{%
  \providecommand\BibTeX{{%
    \normalfont B\kern-0.5em{\scshape i\kern-0.25em b}\kern-0.8em\TeX}}}

\usepackage{makecell} 
\usepackage[font=normalsize,skip=0pt]{subcaption}
\usepackage[subtle]{savetrees}
\usepackage[explicit,noindentafter]{titlesec}
\titlespacing{\section}{0pt}{\parskip}{-\parskip}
\titlespacing{\subsection}{0pt}{\parskip}{-\parskip}
\titlespacing{\subsubsection}{0pt}{\parskip}{-\parskip}
\titlespacing*{\paragraph}{0pt}{5pt}{5pt}
\usepackage{enumitem}
\setlist[itemize]{noitemsep, topsep=0pt}
\setlist[enumerate]{after=\vspace{0pt},noitemsep, topsep=0pt}

\def\compactify{\itemsep=0pt \topsep=0pt \partopsep=0pt \parsep=0pt}
\let\latexusecounter=\usecounter

\definecolor{Red}{rgb}{0.8,0,0}

\newcommand{\system}[0]{Ekiben}

\usepackage{siunitx}
\usepackage{hyperref}
\begin{document}

\title{\LARGE Agile Development of Linux Schedulers with \system{}
\vspace{-25mm}
}

\author{Samantha Miller}
\affiliation{
  \institution{University of Washington}
  \country{USA}
}

\author{Anirudh Kumar}
\affiliation{
  \institution{University of Washington}
  \country{USA}
}

\author{Tanay Vakharia}
\affiliation{
  \institution{University of Washington}
  \country{USA}
}
\author{Danyang Zhuo}
\affiliation{%
  \institution{Duke University}
  \country{USA}
}

\author{Ang Chen}
\affiliation{%
  \institution{Rice University}
  \country{USA}
}

\author{Tom Anderson}
\affiliation{%
  \institution{University of Washington}
  \country{USA}
}


\begin{abstract}
Kernel task scheduling is important for application performance, adaptability to new hardware, and complex user requirements.
However, developing, testing, and debugging new scheduling algorithms in Linux, the most widely used cloud operating system, is slow and difficult.
We developed \system{}, a framework for high velocity development of Linux kernel schedulers.
\system{} schedulers are written in safe Rust, and the system supports live upgrade of new scheduling policies into the kernel, userspace debugging, and bidirectional communication with applications.
A scheduler implemented with \system{} achieved near identical performance (within 1\% on average) to the default Linux scheduler CFS on a wide range of benchmarks. \system{} is also able to support a range of research schedulers, specifically the Shinjuku scheduler, a locality aware scheduler, and the Arachne core arbiter, with good performance.
\end{abstract}



\maketitle
\pagestyle{plain}

\setcopyright{none}

\vspace{-3mm}
\section{Introduction}

Kernel scheduler behavior is central to application performance, adaptability to new hardware, and complex user requirements.
Many applications have short, latency sensitive tasks and bursty workloads, where
suboptimal kernel decisions can lead to high tail latency and poor overall job performance~\cite{shinjuku,shenango,persephone}.
Heterogeneous hardware, such as non-uniform and tiered memory or accelerators, increases the complexity of scheduling decisions~\cite{popcorn,helios}.
Energy use is also increasingly important; neither underloading nor fully loading a server provides peak server energy efficiency~\cite{warehouse-scale,nest}.
Additionally, applications may have information that can help improve scheduling decisions, such as workload characteristics or hardware preferences, but in most kernels the scheduler is oblivious to user preferences beyond simple priorities and hand coded placement~\cite{amd-sched,cerebros}.

Efficiently supporting these changes will require new scheduler designs or new features in existing schedulers.
Although kernel schedulers could theoretically be adapted to handle these new demands,
developing and testing new kernel schedulers can be difficult and time consuming.
Kernel code is difficult to write correctly and debug and slow to deploy. There are only three mainline schedulers implemented in the Linux kernel, the most widely used cloud operating system.

To address this, some researchers have used kernel bypass to implement new schedulers~\cite{shinjuku,shenango,arachne,persephone}.
This approach increases development velocity by removing the need to recompile the kernel and providing access to userspace debugging tools.
However, it interferes with resource sharing between the scheduler and the rest of the system and complicates deployment and maintenance, limiting the potential reach of the research~\cite{interposition}.

GhOSt~\cite{ghost} aims to provide a general purpose deployable solution for userspace schedulers in Linux.
GhOSt uses an upcall approach where scheduling policy decisions are made by a userspace scheduler while the mechanism remains in the kernel. GhOSt schedulers can be implemented in small amounts of userspace code and redeployed easily, but each scheduling decision requires scheduling the userspace scheduler. To hide latency,
ghOSt uses an asynchronous model where the kernel can continue to take interrupts and make scheduling decisions while the userspace scheduler runs. This means the scheduling decisions may be out of date.

Other kernel subsystems, particularly networking, can use eBPF~\cite{ebpf} for high velocity kernel development, and ghOSt implemented support for scheduler eBPF hooks.
Using eBPF, users can load programs to customize kernel scheduler code, provided that the structure of the scheduler does not change.
It is difficult to implement large or complicated code, such as an entire scheduler, using eBPF. For example, the default Linux scheduler, the Completely Fair Scheduler, is over 6,000 lines of code.
Additionally, the eBPF trust model is not a good match for scheduling. eBPF considers loaded programs potentially malicious and verifies that they will not disrupt kernel execution. It is not clear how to implement this for scheduling since bad scheduling policy decisions can violate eBPF's safety requirements.

Previously, Bento~\cite{bento} has shown that we can have high development velocity and high performance for Linux kernel file systems by writing file systems in safe Rust and supporting live upgrade and userspace debugging. However, Bento sits behind the file page cache, reducing the cost of interposition. It is not clear if Bento's design is low enough latency for scheduling.
Additionally, Bento does not address logical correctness errors. In the file system, non-memory safety correctness errors will generally affect only the applications using the file system, while correctness errors in the scheduler code can crash the kernel.

Inspired by Bento, we built a framework called \system{}. The goal of \system{} is to enable high velocity development and deployment of high performance schedulers in the Linux kernel. \system{} schedulers are fast to write and debug and easy to test with seamless resource sharing with the rest of the kernel.
\system{} supports schedulers running in the Linux kernel for wide deployability, but our approach is not restricted to the Linux kernel.
A new \system{} scheduler is written in safe Rust against a clean interface, making it less likely to introduce bugs.
\system{} enables dynamic update of scheduling code in a live kernel without rebooting, with a pause of only 10$\mu$s.
Using a record and replay system, scheduling policies can be debugged using userspace tools. Since \system{} schedulers are implemented in the kernel, they can coordinate easily with other kernel schedulers, such as passing cores between schedulers or applications.

We used \system{} to implement several different scheduling algorithms to demonstrate its flexibility in supporting a variety of scheduler designs and to understand the overhead and performance of \system{} compared to native implementations of the same schedulers.
We implemented a weighted fair queuing scheduler and evaluated it against Linux's default scheduler CFS. Despite our scheduler being much simpler than CFS, and including the \system{} framework overhead, it achieves an average of only 0.74\% slowdown across 36 application benchmarks, with a maximum slowdown of 8.57\%. 
We also implemented the Shinjuku~\cite{shinjuku} scheduler, a locality aware scheduler that co-locates tasks of the same user-defined class, and the core arbiter from the Arachne multilevel thread scheduler.
The \system{} Shinjuku and Arachne schedulers performed competitively with native implementations, and the locality aware scheduler showed the potential to provide significant performance benefit. Two of the schedulers were implemented by undergraduates with no prior Linux kernel programming experience. The code will be made publicly available.

\section{Motivation and Approach}



Linux includes three schedulers: a real time scheduler, an earliest deadline first scheduler, and the Completely Fair Scheduler (CFS). CFS is the default and implements a version of weighted fair queuing.
These schedulers are all quite large and complex; CFS is over 6000 lines of code, and even the simpler real time and deadline schedulers are over 1500 lines of code.
This complexity has led to a number of bugs, particularly performance bugs due to a complicated load balancing mechanism~\cite{wasted-cores}.

Although CFS's weighted fair queuing algorithm works well for many of the tasks run on desktop or server machines. Other schedulers can have advantages in some cases. With more application knowledge, more optimal decisions are possible. For example, for workloads composed of many very small tasks, shorter time slices can lower average job completion time~\cite{shinjuku}. Multilevel scheduling can give better performance isolation and flexibility by allowing applications to define their own policies based on their workloads~\cite{arachne}. 
Nest~\cite{nest} improves energy efficiency for jobs with fewer tasks than cores by reusing warm cores rather than spreading tasks across many cold cores.
Because these schedulers do not need to work well in all circumstances, they can potentially be much smaller and simpler than CFS.
Nevertheless, uptake into Linux has been slow.

We propose \system{}, a framework for high velocity development of Linux kernel schedulers. \system{}'s trust model assumes that scheduler developers are trusted but clumsy; they are not trying to break the kernel but may accidentally introduce bugs. Our overall goal is to allow non-expert programmers to be able to successfully design, implement, debug, and deploy new schedulers.

There are several challenges to achieving high development velocity for Linux kernel schedulers:

\begin{itemize}
    \item Buggy code: The Linux kernel is a monolith of complicated C code, causing bugs to be common. The lack of modularity in the Linux kernel and the potentially large consequences of kernel bugs force developers to code slowly and carefully. In \system{}, we rely on safe language features. By enabling schedulers implemented entirely in safe Rust and by introducing a novel application of the type system to check for scheduling correctness errors, we eliminate whole classes of bugs without introducing significant performance overhead or overly limiting scheduler designs. 
    \item Limited interaction: Some scheduler designs, such as two level schedulers that require coordination with user threads, are difficult to implement in Linux because they require interaction with the userspace tasks that Linux does not support. By allowing generic, scheduler-defined interactions between the scheduler and userspace tasks, \system{} can support scheduler designs that cannot be implemented in Linux today, such as those that require application specific hints or two level schedulers. 
    \item Disruptive upgrade: Current Linux schedulers are compiled into the kernel source, so deploying a new scheduler requires recompiling the kernel and rebooting the machine. \system{} schedulers can be upgraded quickly, without rebooting the machine and with only a short period of service downtime.
    \item Slow debugging: The kernel does not have access to debugging tools, such as those commonly used in userspace. We support record and replay debugging in \system{}. To diagnose bugs not caught by the \system{} framework, developers can record the calls between the kernel and the scheduler and replay later entirely at userspace.
    \item Resource sharing: Schedulers should be able to share resources with the rest of the system. In Linux, different applications can use different schedulers, sharing cores and cycles between the schedulers. \system{} schedulers are implemented in the kernel to enable fine grained core sharing across applications and schedulers. 
\end{itemize}







\section{\system{}}

\begin{figure*}
    \centering
    \includegraphics[width=0.8\textwidth,trim={1cm 2cm 2cm 2.5cm},clip]{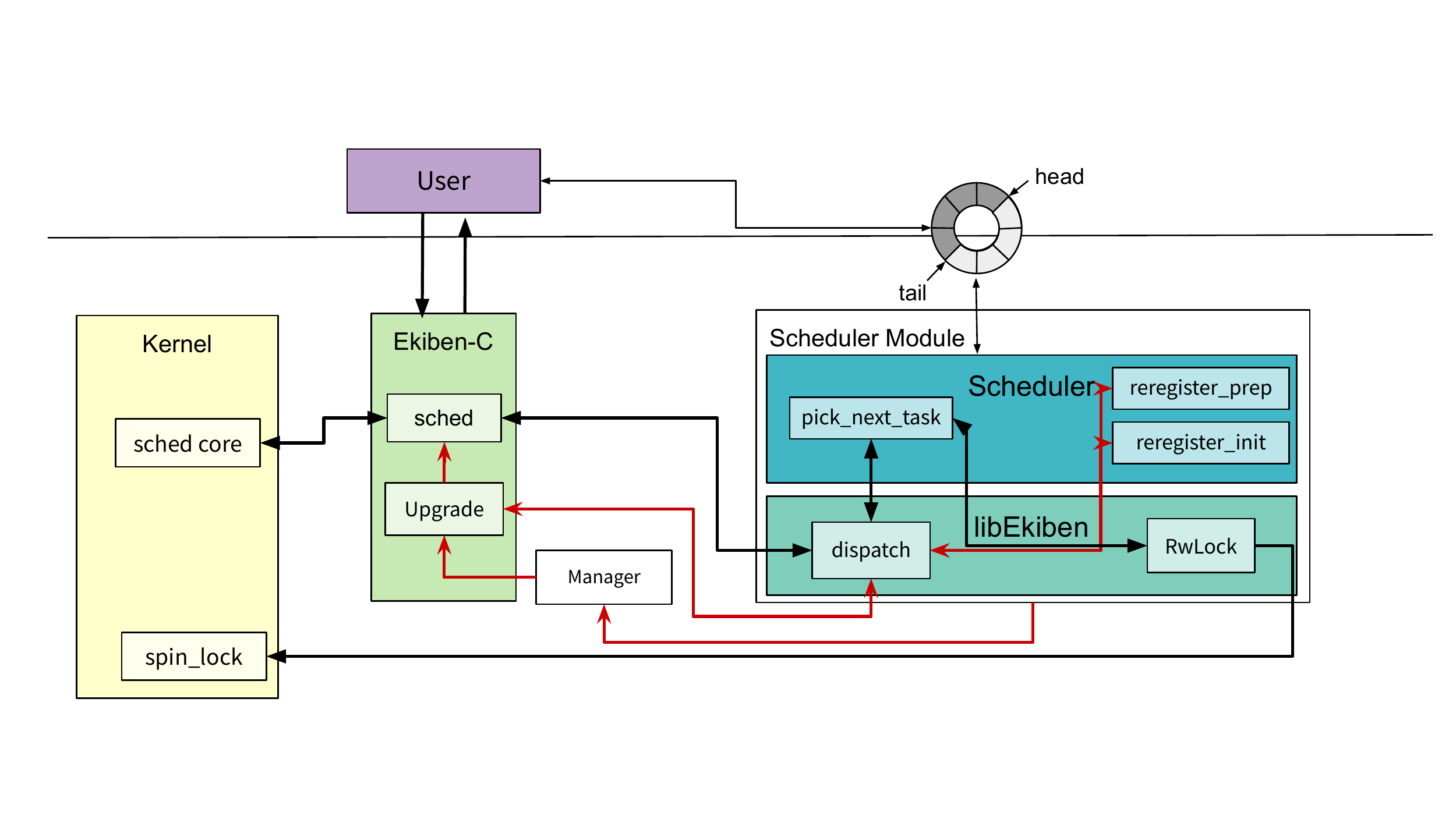}
    \caption[A high level diagram Ekiben.]{A high level diagram \system{}. \system{}-C is written in C and compiled into the kernel. It interacts with lib\system{}, a Rust library that is compiled with the scheduler code into the dynamically loaded scheduler module. The black lines represent code pathways during normal execution. Red lines represent module insertion and upgrade.\vspace{-4mm}}
    \label{fig:ekiben}
\end{figure*}

The high level overview of \system{} is shown in \autoref{fig:ekiben}. \system{} is composed of two major components, \system{}-C and lib\system{}.



Part of \system{}, \system{}-C is implemented in C and compiled into the Linux kernel. It interfaces directly with the core scheduling code and the kernel scheduling data structures.
\system{}-C handles registering, deregistering, and upgrading schedulers and sets up and manages infrastructure for communication channels between userspace and the kernel scheduler and the record and replay system.
It handles the unsafe work that is required for scheduling on behalf of \system{} schedulers, such as performing state updates to kernel \texttt{task\_struct} data structure, managing interactions with the kernel run-queues when adding or moving tasks, and manipulating raw pointers to read and pass data to the scheduler module.
\system{}-C also translates the calls from the core scheduling code into calls that the scheduler can implement safely by ensuring that all data passed to the scheduler can be safely accessed. \system{} schedulers do not directly manipulate kernel state or run-queues.


The other component, lib\system{}, is a Rust library that is compiled with the scheduler code into a module that is dynamically loaded into the kernel. This library provides safe interfaces so the scheduler code can access the kernel and implements functions for loading and managing the scheduler. It contains some unsafe Rust because it must handle interactions with the C code in \system{}-C, which is inherently unsafe. Each scheduler is written entirely in safe Rust and only needs to provide the logic for the scheduling algorithm. The scheduler module is not sandboxed further; once it is loaded into the kernel, it runs like any other kernel code.

When the scheduler module is loaded, lib\system{} calls \system{}-C to register the newly available scheduler.
This registers the ID of the scheduler being loaded and a processing function in lib\system{} for parsing calls from \system{}.
User tasks can switch to using the new scheduler using its defined ID value.
During regular operation, \system{}-C processes calls from the core scheduler code for these tasks, forwarding the calls to the processing function in lib\system{} and managing updates to kernel data structures, such as the CPU's run queue.
When the module is unloaded, lib\system{} similarly unregisters the scheduler with \system{}-C, and no new tasks can be attached to the scheduler.

\subsection{Safe Interfaces}
\system{} provides schedulers with safe interfaces, both the interface that schedulers are required to implement and the interfaces for schedulers to access kernel functionality, such as locks and timers. With safe interfaces, \system{} schedulers can be implemented entirely in safe Rust, preventing whole classes of bugs and reducing time spent debugging new schedulers.
\system{}-C and lib\system{} work together to provide safe interfaces for the scheduler code. 

\system{}'s threat model assumes that the developer is well meaning and knowledgeable but makes occasional mistakes. \system{} aims to help this developer prevent these mistakes by catching low-level errors that do not depend on the scheduler behavior, such as NULL pointer dereferences and data races, at compile time. \system{} does not aim to prevent all bugs, and bugs that depend on the scheduler's semantic behavior can remain uncaught. For example, schedulers implemented with \system{} can deadlock, lose tasks, and violate work conservation.
We attempt to catch as many of these bugs as we can at runtime, but cannot guarantee that all instances are caught.

\system{}-C takes the interface defined by the core scheduler code and translates it into an interface based on message passing. 
Like the core scheduler code, this is a synchronous interface; it consists of function calls where the caller waits for the the callee to return before progressing.
It is not necessary for safety that the interface be based on message passing, but it helps enforce certain safety properties, such as no shared pointers, preventing memory bugs across the interface.
\system{}-C handles direct interactions with kernel data structures, such as pulling information like the runtime of the task or the current CPU of the task. This information is placed into per-function type "message" data structures that are passed to the registered processing function in lib\system{}.

The processing function in lib\system{} parses each "message" to determine which scheduler function is being invoked and handles the unsafety of interacting directly with C code through the Rust FFI (Foreign Function Interface) layer. The processing function pulls the fields from the "message" and passes them to the scheduler function being called. If the function returns a value, lib\system{} writes that value back into the "message" data structure to return the value to \system{}-C.

To enforce that the scheduler code implements the required behavior, lib\system{} provides a Rust trait for scheduler types. This trait defines the functions that a scheduler module must implement to be loadable as an \system{} scheduler. A trait defines a set of functionality that a type must have to implement the trait and can provide default implementations for shared behavior on those types. Traits can be used as bounds on function arguments, and any type that implements a trait can be used where the trait is called for.

\begin{table}[ht!]
    \centering
    \begin{tabular}{l|l}
    \Xhline{4\arrayrulewidth}
        \textbf{API Function} & \textbf{Description}\\
        \hline
        \texttt{pick\_next\_task}
            & Pick the next task for the CPU. \\
        \texttt{pnt\_err} & Could not schedule chosen task. \\
        \texttt{task\_new} & There is a new task. \\
        \texttt{task\_wakeup} & A task woke up. \\
        \texttt{task\_blocked} & A task blocked. \\
        \texttt{task\_dead} & A task died. \\
        \texttt{task\_tick} & A timer has triggered. \\
        \texttt{select\_task\_rq} & Choose the CPU for a task. \\
        \texttt{migrate\_task\_rq} & A task is moving CPUs. \\
        \texttt{balance} & Rebalance tasks onto CPU. \\
        \texttt{balance\_err} & Could not move the chosen task. \\
        \texttt{register\_queue} &
            Register a user to kernel hint queue. \\
        \texttt{enter\_queue} & Check the user hint queue.\\
        \texttt{unregister\_queue} &
            Unregister the user to kernel queue.\\
        \texttt{parse\_hint} & Synchronously parse hint.\\
        \texttt{reregister\_prep} & Prepare to be unloaded in upgrade. \\
        \texttt{reregister\_init} & Initialize in upgrade. \\
        \Xhline{4\arrayrulewidth}
    \end{tabular}
    \caption[The API of the EkibenScheduler Trait.]{A selection of the API of the \system{}Scheduler Trait. This is the API that a scheduler module should implement to be loadable as an \system{} scheduler.\vspace{-6mm}}
    \label{tab:ekiben_api}
\end{table}

The \system{}Scheduler trait (shown in \autoref{tab:ekiben_api}) specifies the functions that a scheduler should provide. 
Most of these functions are very similar to the functions defined by the core scheduling code to implement a Linux scheduler, with some notable differences.

The core function for a scheduler is \texttt{pick\_next\_task} which tells the core kernel scheduler code which task should be run next. Other important functions are \texttt{task\_new}, \texttt{task\_wakeup} and similar functions for tracking task state, \texttt{migrate\_task\_rq} for moving tasks between cores, \texttt{balance} for telling the core scheduler to move tasks to rebalance load, and error functions to return error values from the kernel.
While these functions are modeled after the Linux scheduler interface they could be adapted to a different operating system.
An \system{} scheduler is only expected to manage its own state in response to these calls; the kernel's core scheduling code decides when to call each function and \system{}-C manages kernel state.

In Linux, schedulers are expected to track their tasks' runtimes using kernel timing functions. In our system, \system{}-C tracks the runtime of tasks on behalf of the scheduler and passes this to the scheduler when the task state changes, such as blocking, waking, and yielding, and to \texttt{pick\_next\_task}. \system{} records this information for deterministic replay.

The \texttt{pick\_next\_task} function in Linux expects the scheduler to choose a task on the CPU's run-queue, and if this expectation is violated, the kernel can crash.
To prevent this bug, we introduce a new type called \texttt{Schedulable} that represents a task and what core it can safely be scheduled on.
When tasks are created, blocked or unblocked, or moved between run-queues, lib\system{} creates a \texttt{Schedulable} data structure to indicate which core's run-queue the task is on and passes ownership of it to the scheduler at the corresponding call.
The scheduler returns the data structure back to \system{} as the return value for \texttt{pick\_next\_task} as proof that the task can be safely scheduled on the core. In lib\system{}, the core in the \texttt{Schedulable} is checked against the core being assigned, and if the check fails, the data structure's ownership is returned to the scheduler using the \texttt{pnt\_err} call.
A \texttt{Schedulable} also cannot be copied or cloned, so the scheduler cannot hold onto an old \texttt{Schedulable} to act as validation after it has returned it to \system{} when moving or running the task. The scheduler must instead receive a new \texttt{Schedulable}, such as through the \texttt{task\_wakeup} call or as the current task in \texttt{pick\_next\_task}.

\system{} cannot always prevent the scheduler from holding onto an invalid \texttt{Schedulable.}
The \texttt{migrate\_task\_rq} function, which moves tasks between cores, passes in a new \texttt{Schedulable} for the new core and requires the scheduler to return the old one so the scheduler will only have validation to run the task on one core. We cannot check at compile time that the scheduler returns the correct \texttt{Schedulable} for the old core, so it is possible for a scheduler to keep the wrong data structure.
Additionally, we cannot require that the scheduler return a \texttt{Schedulable} in \texttt{task\_blocked} or \texttt{task\_dead} because these functions are sometimes called when the task is not schedulable and the scheduler has nothing to return.

In lib\system{}, we also provide a safe interface for receiving hints from userspace (\autoref{sec:hints}).

\subsection{Live Upgrade}

We support live upgrade of schedulers in \system{}. Live upgrade allows an upgraded version of a scheduler to replace an older version of the same scheduler without rebooting the machine or killing any of the tasks using the scheduler. Because the scheduling code is running throughout the upgrade, we must ensure that any state maintained by the scheduler, such as task runtimes and which task is runnable on which core, remains consistent across the upgrade and is available to the new scheduler after the upgrade.

We ensure consistency of the state by quiescing the scheduler module during the upgrade. Since the scheduler module is only invoked from \system{}-C, we know that the state of the scheduler will remain consistent as long as all calls from \system{}-C are completed before the upgrade begins and any new calls wait until the upgrade is completed. We implement this using a simple per-scheduler read-write lock. Non-upgrade calls into the scheduler module acquire the lock in read mode, allowing multiple concurrent calls into the scheduler module. When an upgrade begins, the lock is acquired in write mode, preventing any of the non-upgrade calls from entering the scheduler module. When the upgrade is finished, the lock is released, and the non-upgrade calls can proceed, now directed to the new version of the scheduler.

After the scheduler module is quiesced, \system{}-C calls into \texttt{reregister\_prep}, informing the scheduler that an upgrade is occurring. The module then performs any necessary maintenance and returns a data structure with any state it wishes to pass to the new version of the scheduler. \system{}-C then calls \texttt{reregister\_init} in the new version of the scheduler, passing the data structure returned from the old scheduler. The new scheduler can then initialize itself based on the provided state, claiming ownership of some or all of the state if it wishes. Because \system{}-C acquired the read-write lock described above in write mode, we know that no other calls can enter either scheduler module during the upgrade, and it is safe for the scheduler to manipulate its internal state.

After ensuring that state remains consistent across the scheduler during an upgrade, the rest of the upgrade is fairly straightforward. \system{}-C swaps pointers so that its internal data structure now points to the new module, and the upgrade completes. Calls to the scheduler will use the new pointer, and so will call into the new scheduler module. The old scheduler module can then be safely unloaded.


\textbf{Limitations.} Because we quiesce the scheduler during an upgrade, there is a period when the scheduler is blocked and cannot accept non-upgrade calls, leading to a short service blackout.
\system{} trusts the scheduler to have short, well-defined code paths so the read locks will be given up quickly and the upgrade can progress. We also trust the scheduler to upgrade and release the write lock quickly.
Another limitation is that the state data structure that the new scheduler expects must be the same as the data structure exported by the old scheduler because the memory is passed directly. This data structure is otherwise completely custom, and the new scheduler can export a different data structure to the next scheduler upgrade.

\subsection{Custom Scheduler Hints}\label{sec:hints}

Correct scheduling decisions often depend on the behavior of the workload being scheduled. Tasks that communicate with each other or operate on the same data benefit from being scheduled on the same NUMA node, or with more recent split last level cache architectures, on cores with the same last level cache~\cite{amd-sched}. On a heterogeneous CPU, some tasks might prefer to be scheduled on certain cores or devices or co-located with other tasks~\cite{helios}.

In addition to sending scheduling hints from userspace to the kernel, it can also be useful for information to flow from the kernel to userspace. For example, in scheduler activations~\cite{activations}, the scheduler provides information about scheduling events to the user-level thread scheduler, such as when cores become available or when user tasks block in the kernel.
Another example could be to utilize machine learning to improve scheduling decisions~\cite{10.5555/3023549.3023573}.

\system{} supports custom scheduler-defined hints, both from userspace to the kernel and vice versa. Each scheduler that supports hints defines data structures indicating the type of hints that it expects to receive and the type of hints it will send to the user. We enforce that these types can be read-shared across the user/kernel boundary without violating memory safety, but otherwise put no restrictions on them.

Queues can be shared across a live upgrade as long as both versions of the scheduler use the same hint data structures. The scheduler passes the queues as part of the shared state during the upgrade. If the next scheduler version will use different hints, old queues must be closed before the upgrade, and new ones can be created after.


\subsection{Record and Replay}

Kernel debugging is notoriously difficult. Many debugging tools are difficult or even impossible to use on the kernel. While \system{} prevents many bugs that would crash the kernel, logic bugs can still exist in the schedulers. We want to provide a mechanism to simplify debugging by allowing the scheduler code to be run and debugged at userspace.


To this end, \system{} implements a record and replay system for scheduling events. When running in record mode, lib\system{} records each call and hint sent to the scheduler. The replay system implements a replacement version of lib\system{} to replay these records to the scheduler, now running in userspace. The exact same scheduler code is run during both record and replay. 
The replay is primarily aimed at understanding the behavior of the recorded scheduler. The trace could be used as input to a modified scheduler, but the policy changes may affect the order of scheduler actions and the scheduler behavior.
Consulting a userspace scheduler synchronously on every operation, like Bento's approach to file system debugging, is infeasible because the scheduler subsystem cannot block waiting for a response from a userspace program that itself needs to be scheduled to run.
Likewise, consulting a userspace scheduler asynchronously, as in ghOSt, can result in different behavior between the kernel and user versions.

\textbf{Record. } Lib\system{} must record the sequence of information provided to the scheduler, both procedure calls into the scheduler and hints from the queue, so they can be replayed exactly as they were originally played in the kernel. We also record responses returned by the scheduler so we can alert the user if the scheduler returns a different result during replay. We do not record or attempt to replay the exact timing of the messages. Where timing is relevant to scheduler decisions, such as the runtime of a task, that information is provided in the message from the kernel and so will be recorded. We assume that the scheduler does not attempt to validate this information or track its own timing.

\system{}'s record and replay system must also handle concurrency. The Linux kernel is multi-threaded, and multiple kernel threads can call into an \system{} scheduler at once, e.g. on different cores. Due to Rust's properties and \system{}'s structure, we know that potential race conditions are protected by locks, and we can identify and record the order of lock acquisitions. As long as locks are acquired in the same order during record and replay and the behavior of the scheduler is deterministic, the results should be the same~\cite{r2}. In lib\system{}, we include recording functionality in the shim wrappers around the kernel lock functions to record lock creation, acquisition and release, along with the address of the lock and the ID of the accessing kernel thread. All other message records are also tagged by the thread ID of the kernel thread calling into the scheduler.

Recording messages in the scheduler stack is non-trivial. In many cases, \system{} is called while the kernel has interrupts disabled. Writing to a file has the potential to sleep, so we cannot write the messages to a log file while in the scheduler context. Even printing to the kernel log must be delayed until out of the scheduler context.

Similarly to userspace hints, we use a ring buffer to solve this. We run a separate userspace task that creates a ring buffer queue shared with \system{}-C. When lib\system{} wants to record a message, it sends it to \system{}-C, which adds the message to the queue. The userspace task consumes messages on the queue and writes them to a file. If the buffer overruns, events may be dropped.

In order to record messages, the userspace record task must be running and the scheduler must have been compiled in record mode. By default, lib\system{} does not record messages.

\textbf{Replay. }
Replay consumes the file created during recording. The replay utility sends the recorded messages directly to the scheduler code in the same order they were called and validates the responses against the recorded ones.

To handle concurrent replay, we ensure that all locks are acquired in the same order in replay as they were during record. First, the replay system analyzes the log and parses out the lists of operations on each lock, using the lock's memory address to differentiate the locks. The locks are then created, passing in the list of acquisitions for each lock. We assume that locks are created in the same order during replay as they were during record and are not deallocated.

To allow for concurrent operations, the replay system starts a thread per recorded message in the record log as it replays the log.
When each replay thread is created, the replay system names it with the ID of the associated kernel thread. When the replay thread attempts to acquire a lock, the lock checks whether it is the next to acquire the lock. If not, the thread is blocked until its turn.

\system{} does not support upgrading the scheduler during the record and replay process.

\section{Implementation}

\subsection{\system{}}

\begin{table}[]
    \centering
    \begin{tabular}{l|l|c|c}
    \Xhline{4\arrayrulewidth}
        \textbf{Component} & \textbf{Lang.} & \textbf{LOC} & \textbf{Unsafe LOC} \\
        \hline
        \system{}-C & C & 2411 & N/A \\
        Scheduler lib\system{} & Rust & 962 & 94 \\
        Other lib\system{} & Rust & 5870 & 2858 \\
        Userspace Record & Rust & 95 & 10 \\
        Replay & Rust & 646 & 0 \\
         \Xhline{4\arrayrulewidth}
    \end{tabular}
    \caption{Lines of code for the \system{} components. The scheduler lib\system{} includes the \system{}Scheduler trait, hint queues, and support for record and live upgrade. Other lib\system{} provides safe access to kernel data structures and functions. Part of Other lib\system{} is borrowed from the analogous library from Bento (libBentoKS).\vspace{-5mm}}
    \label{tab:loc}
\end{table}

The lines of code for \system{} are shown in \autoref{tab:loc}. \system{} is implemented in Linux 5.11 and is based on the kernel component from ghOSt~\cite{ghost}. The scheduler lib\system{} includes the \system{}Scheduler trait, hint queues, and support for record and live upgrade. Other lib\system{} provides safe access to kernel data structures and functions and general support for Rust kernel programming.


\subsection{Schedulers}
To demonstrate \system{}'s flexibility, we implemented a selection of schedulers: a weighted fair queuing scheduler based on Linux CFS, a specialized research scheduler based on Shinjuku, and schedulers that utilize the userspace hinting and bidirectional communication features of \system{} for locality aware scheduling and two level scheduling. As baselines, we use the default Linux CFS and the ghOSt~\cite{ghost}


\subsubsection{CFS and Weighted Fair Queuing. }
\textbf{ }
CFS uses per-core run-queues, meaning it first assigns tasks to cores, and then chooses the next task for the core from among the assigned tasks.
On each core, CFS implements a version of weighted fair queuing, dividing CPU time proportionally between groups of tasks, and then within each group, while respecting priority.
It uses a calculation called \textit{vruntime} to track which group/task to choose next, choosing the group/task with the lowest weighted accumulated runtime.
The \textit{vruntime} is calculated based on the task's runtime, modified by its priority; tasks with higher priority accrue \textit{vruntime} slower.
To prevent sleeping tasks from accruing a large \textit{vruntime} debt and therefore running for too long after they wake, newly woken tasks receive the maximum of their old \textit{vruntime} and the \textit{vruntime} of the task with the lowest \textit{vruntime} in the run-queue minus a several millisecond threshold.
If a newly woken task has a smaller \textit{vruntime} than the current task, it preempts the current task when a system timer ticks.
Otherwise, task switches occur only after a task has run for its allocated time slice.
In order to prevent starvation, CFS attempts to run every task at least once per time period, where the period depends on the number of tasks, with a minimum of 6ms. 
When tasks are created or woken, the length of the period adjusts to include the new task. New tasks are run at the end of the period, but recently woken tasks can be run earlier.

CFS places running tasks onto cores and moves tasks between cores to achieve better performance.
By default, CFS will attempt to even out the amount of work per core, based on information such as
the amount of time the core is idle or overloaded, the priority of the tasks on the core, the capacity of the cores on the machine, and the preferences of the tasks.
In certain cases, CFS will co-locate tasks on specific cores, such as if it is required to by the user.
CFS attempts to place tasks so that newly woken tasks can be scheduled promptly. When a task is woken or a core becomes idle, CFS will move tasks so the newly idle cores are used.
CFS rebalances task placement every 1-10ms, depending on the configuration, or when cores become newly idle.
CFS first tries to move tasks to cores within the same NUMA node, and will not balance tasks across NUMA nodes unless there are more than a threshold more tasks running on the busier NUMA node.

\paragraph{\system{} Weighted Fair Queuing.}
To evaluate the overhead of \system{}, we implemented our own scheduler based on CFS.
Our version does not provide the full complexity of the algorithm; instead, we are interested in showing the overhead of our approach on benchmarks relative to CFS.
We compute \textit{vruntime} for per-core time slices but use a much simpler method for determining task placement. If a core is about to become idle and another core had a waiting task, our scheduler steals waiting work from the core with the longest queue of tasks. Otherwise, our scheduler does not rebalance tasks. We found this compromise allowed our scheduler to achieve good performance on a wide array of benchmarks with much less complexity - 646 lines of code versus 6247 for CFS. 


\subsubsection{GhOSt and the Shinjuku Scheduler. } \textbf{ }GhOSt~\cite{ghost} is a research framework for userspace schedulers. GhOSt implements a trampoline approach; calls from the core scheduler code are forwarded to the userspace scheduler, which sends its scheduling decisions to the kernel. GhOSt uses an asynchronous message passing model, i.e. the core scheduler code does not wait for the userspace scheduler to respond before choosing what to run.
We evaluate microbenchmarks against two ghOSt schedulers, the per-CPU FIFO scheduler, which runs per-core schedulers that manage tasks assigned to that core, and the SOL scheduler, a latency-optimized FIFO scheduler that manages all cores, running the scheduler on a separate core.
We also implement a version of the Shinjuku~\cite{shinjuku} scheduler and evaluate it against ghOSt's version of the same scheduler.

The Shinjuku scheduler~\cite{shinjuku} achieves low latency for workloads with short, high priority tasks and longer, low priority tasks with an efficient version of shortest task first. Shinjuku uses centralized first-come-first-serve scheduling. After each task has run for 5 to 15$\mu$s Shinjuku preempts it, placing it at the back of the queue. However, Shinjuku does not run in Linux, instead depending on the Dune~\cite{dune} operating system. Unlike Linux, Dune applies a single scheduling algorithm for all tasks on the machine.

To evaluate \system{}'s ability to support a specialized scheduler for Shinjuku-style workloads, we implemented a scheduler based on the Shinjuku scheduler using \system{}. Our scheduler implements an approximation of a first-come-first-serve queue of tasks with fast preemption across the multiple kernel run-queues. Our preemption slice is 10$\mu$s instead of 5$\mu$s to prevent overloading the scheduler. This scheduler was implemented in 285 lines of code.

\subsubsection{Locality Aware Scheduler. }
\textbf{ } We also implemented a locality aware scheduler using \system{} that co-locates tasks that communicate heavily with each other or benefit from cache sharing.
This scheduler uses \system{}'s userspace hinting mechanism to inform the scheduler about which tasks to co-locate.
The application sends the ID of each newly created thread and a locality value to indicate which tasks should be co-located. 
Unlike Linux's \textit{taskset} cgroup, these hints do not need to specify the core for each task, only its colocation, which the scheduler can ignore if non-optimal, such as when there are too many tasks on a given core. This scheduler was implemented in 203 lines.

\subsubsection{Two-level Scheduler. }
\textbf{ }The Arachne user-level scheduler provides two-level thread management: applications request cores and manage user-level threads on the assigned cores~\cite{arachne,activations}.

In Arachne, both the core arbiter and the runtime are implemented in userspace. The core arbiter relies on Linux's \texttt{cpuset} mechanism to manage core assignments. The runtime sends messages to the core arbiter over a socket, and the core arbiter either responds on the socket or uses a shared memory page. This socket allows the runtime to manually block if a core is not available. 

We reimplemented the Arachne core arbiter as a kernel scheduler using \system{}. This scheduler uses \system{}'s bidirectional userspace hints.
We use the user-to-kernel queue to send core requests to the \system{} core arbiter; we use the kernel-to-userspace queue for core reclamation requests. The \system{} core arbiter executes the same decisions as the Arachne core arbiter, but uses standard kernel scheduling mechanisms for assigning, moving, and blocking user scheduler activations rather than relying on \texttt{cpuset} and sockets. The \system{} version of the core arbiter is implemented in 579 lines of code. We compare this scheduler against both CFS and unmodified Arachne.

\section{Evaluation}
In our evaluation, we test \system{}'s ability to meet our goal of high velocity development for a wide variety of high performance schedulers with minimal runtime overhead. In \autoref{discussion}, we discuss the development experience of \system{}. In this section, we evaluate the performance of the baseline and \system{} schedulers. We evaluate whether these schedulers can achieve equivalent performance to native implementations on microbenchmarks and application workloads. For research schedulers, we evaluate the scheduler against using benchmarks from the original paper.

\begin{table*}[ht!]
    \centering
    \begin{tabular}{l|c|c|c||c|c|c|c}
        \Xhline{4\arrayrulewidth}
        \textbf{Message Latency ($\mu$s)} & \textbf{CFS} & \textbf{GhOSt SOL} & \textbf{GhOSt FIFO} & \textbf{WFQ} & \textbf{Shinjuku} & \textbf{Locality} & \textbf{Arachne}\\
        \hline
        \normalsize
        One Core & 3.0 & 6.0 & 9.1 & 3.6 & 4.0 & 3.5 & 0.1 \\
        Two Cores & 3.6 & 5.8 & 7.0 & 4.0 & 4.4 & 3.9 & 0.2 \\
        \Xhline{4\arrayrulewidth}
    \end{tabular}
    \caption{Scheduler latency for the \texttt{perf bench sched pipe} benchmark in $\mu$s per wakeup. \system{} adds around 0.6$\mu$s of latency over CFS in the worst case and outperforms the ghOSt schedulers due to the synchronous nature of the benchmark. \vspace{-3mm} }
    \label{tab:perf_results}
\end{table*}
\subsection{Setting}
Benchmarks were performed on either an 8 core, one-socket machine with an Intel i7-9700 CPU running at 3.00GHz or (for scalability tests) an 80 core, two-socket machine with Intel Xeon Gold 6138 CPUs running at 2.00GHz.

\subsection{Microbenchmarks}
We used microbenchmarks to evaluate the latency and scalability imposed by the \system{} framework. We ran the same microbenchmarks on all the schedulers we implemented.


\begin{table*}[h!]
    \centering
    \begin{tabular}{l|l|c|c|c||c|c|c|c}
        \Xhline{4\arrayrulewidth}
        \multicolumn{2}{l|}{\textbf{Worker Threads}} & \textbf{CFS} & \textbf{GhOSt SOL} & \textbf{GhOSt FIFO} & \textbf{WFQ} & \textbf{Shinjuku} & \textbf{Locality} & \textbf{Arachne}\\
        \hline
        \normalsize
        2 Tasks ($\mu$s) & 50th & 74 & 66 & 101 & 78 & 79 & 80 & 1 \\
        & 99th & 101 & 132 & 170 & 104 & 109 & 105 & 1 \\
        \hline
        40 Tasks ($\mu$s) & 50th & 139 & 192 & 152 & 170 & 168 & 175 & 1 \\
        & 99th & 320 & 1354 & 1806 & 323 & 307 & 324 & 1\\
        \Xhline{4\arrayrulewidth}
    \end{tabular}
    \caption{Schbench benchmark with two message threads and 2 and 40 worker threads per message thread. Thread wakeup latencies are measured in $\mu$s.\vspace{-5mm} }
    \label{tab:schbench_results}
\end{table*}
\textbf{Latency. }
\autoref{tab:perf_results} shows the results of the \texttt{perf bench sched pipe} benchmark, averaged over three runs of the benchmark. This benchmark starts two tasks that send 1 million messages back and forth using the \texttt{pipe} system call. 
After each message, the sending task sleeps until the other task responds.
By default, all schedulers put the two tasks on different cores. We also ran the benchmarks forcing both tasks to be on the same core.

Compared to CFS, our \system{} WFQ scheduler adds 0.4$\mu$s (0.6$\mu$s) of latency per message, for the two (one) core case. This represents a 12\% to 20\% overhead on this benchmark. This overhead could possibly be reduced with further optimization. Our version of the Shinjuku scheduler has slightly higher overhead because it starts a reschedule timer on every operation, while CFS and our WFQ scheduler only start a reschedule timer when multiple tasks are present. The locality aware scheduler is slightly faster because it is simpler. The \system{} version of Arachne is much faster than the others because it uses userspace threads instead of processes for blocking and waking threads.

The GhOSt schedulers perform significantly worse than both CFS and \system{}. The per-CPU FIFO scheduler performs worse when both tasks are placed on the same core because they are sharing the core with the ghOSt userspace scheduler. On every schedule operation, the scheduler first must be scheduled and run on the core.




\textbf{Scalability. }
To measure the scalability of \system{}, we evaluate the tail latency of task schedules when there are a large number of tasks using the schbench benchmark. This benchmark starts a number of message threads and worker threads. Each message thread and its worker threads send messages back and forth. Schbench reports the median and 99\% tail latency of task schedules throughout the benchmark. The tested configurations use 2 message threads and 2 or 40 worker threads, resulting in a maximum of 80 worker threads, the same as the number of cores on the machine. When the number of worker threads is larger than the number of cores, the scheduling latency is influenced by waiting time more than scheduler performance. We use a 5s warmup time for each run.

The results are shown in \autoref{tab:schbench_results}. CFS and the \system{} WFQ scheduler showed similar results except for the median with 40 threads.
The \system{} version of Arachne has lower latency than the other schedulers because of its userspace mechanisms for blocking and waking threads.

\subsection{WFQ Scheduler Applications}

\begin{table}[h!]
    \centering
    \begin{tabular}{l|S[table-format=6.2]|S[table-format=6.2]|S[table-format=2.3]}
        \Xhline{4\arrayrulewidth}
        \textbf{Benchmark} & \textbf{CFS} & \textbf{WFQ} & \\
        \hline
        \textbf{NAS Benchmarks} & & & \\
        BT (total Mops/s) & 26669.1 & 26682.3 & -0.05\% \\
        CG (total Mops/s) & 4535.8 & 4475.7 & 1.32\% \\
        EP (total Mops/s) & 487.9 & 491.9 & -0.83\% \\
        FT (total Mops/s) & 14886.8 & 14716.5 & 1.14\% \\
        IS (total Mops/s) & 1297.4 & 1284.9 & 0.96\% \\
        LU (total Mops/s) & 30469.4 & 29811.4 & 2.16\% \\
        MG (total Mops/s) & 8601.4 & 8535.9 & 0.76\% \\
        SP (total Mops/s) & 11797.0 & 11705.6 & 0.78\% \\
        UA (total Mops/s) & 73.8 & 73.1 & 0.87\% \\
        \hline
        \textbf{Phoronix Multicore} & & & \\
        Arrayfire, 1 (GFLOPS) & 812.98 & 820.29 & -0.90\% \\
        Arrayfire, 2 (ms) & 26.72 & 26.71 & -0.04\% \\
        Cassandra, 1 (Op/s) & 55100 & 50573 & 8.22\% \\
        ASKAP, 4 (Iter/s) & 161.46 & 161.12 & 0.22\% \\
        Cpuminer, 2 (kH/s) & 51363 & 51390 & -0.05\% \\
        Cpuminer, 3 (kH/s) & 35667 & 35390 & 0.78\% \\
        Cpuminer, 4 (kH/s) & 9499.87 & 9494.90 & 0.05\% \\
        Cpuminer, 6 (kH/s) & 258100 & 261667 & -1.38\% \\
        Cpuminer, 11 (kH/s) & 29400 & 29323 & 0.26\% \\
        Ffmpeg, 1, 1 (s) & 23.98 & 24.73 & 3.13\% \\
        \makecell[l]{Graphics-Magick, 4 \\ \qquad (Iter/m)} & 781 & 779 & 0.26\% \\
        OIDN, 1 (Images/s) & 0.31 & 0.30 & 3.23\% \\
        OIDN, 2 (Images/s) & 0.31 & 0.30 & 3.23\% \\
        OIDN, 3 (Images/s) & 0.15 & 0.15 & 0\% \\
        Rodina, 3 (s) & 159.32 & 160.00 & 0.43\% \\
        Zstd, 2 (MB/s) & 856.1 & 782.7 & 8.57\% \\
        Zstd, 4 (MB/s) & 153.1 & 165.4 & -8.03\% \\
        AVIFEnc, 4 (s) & 14.94 & 15.33 & 2.62\% \\
        Libgav1, 1 (FPS) & 262.95 & 261.21 & 0.66\% \\
        Libgav1, 2 (FPS) & 67.28 & 66.58 & 1.04\% \\
        Libgav1, 3 (FPS) & 222.70 & 216.51 & 2.78\% \\
        Libgav1, 4 (FPS) & 64.10 & 63.54 & 0.87\% \\
        OneDNN, 4, 1 (ms) & 4.26 & 4.18 & -1.85\% \\
        OneDNN, 5, 1 (ms) & 9.71 & 9.10 & -6.31\% \\
        OneDNN, 7, 1 (ms) & 4166.31 & 4164.74 & -0.04\% \\
        OneDNN, 7, 2 (ms) & 4166.40 & 4161.15 & -0.13\% \\
        OneDNN, 7, 3 (ms) & 4164.25 & 4163.34 & -0.02\% \\
        \Xhline{4\arrayrulewidth}
    \end{tabular}
    \caption{Performance comparison of Linux CFS and \system{} WFQ on the NAS Parallel Benchmarks and a selection of the Phoronix Multicore benchmarks (version 10.8). Full names are provided in the appendix. The maximum slowdown is 8.57\%. The geometric mean over all benchmarks is 0.74\%.\vspace{-5mm}}
    \label{tab:app_results}
\end{table}

To evaluate whether our WFQ scheduler performs competitively with CFS, we ran multithreaded application benchmarks using benchmark suites that covered a wide range of use case patterns. We ran benchmarks from the NAS Parallel Benchmark suite~\cite{nas} and the Phoronix Multi-core Test Suite~\cite{phoronix}. The NAS Parallel Benchmark suite from NASA is a set of benchmarks to evaluate parallel performance on supercomputers.
We ran nine of the ten NAS Benchmarks, excluding the DC benchmark that targets computational grids. We used the "C" benchmark size, the largest standard benchmark size.
The Phoronix Multi-Core Test Suite contains a very large collection of multithreaded application benchmarks. There are over 90 applications, and many have multiple workloads. We report the same collection of 27 benchmarks as reported in Nest~\cite{nest}. Phoronix runs each benchmark three times unless the standard deviation is greater than 5\%, in which case it will rerun the benchmark until the standard deviation is low enough, up to a maximum of 15 times. The results for both benchmark suites are in \autoref{tab:app_results}.

The NAS benchmarks show very little performance difference between the CFS and \system{} WFQ schedulers, with a maximum of 2.16\% difference on the LU benchmark. The NAS benchmarks all start one task per core, so this is expected behavior.
Performance on the Phoronix benchmarks is also quite similar for both schedulers across the benchmarks. The largest slowdowns were 8.22\% and 8.57\% on the Cassandra Writes benchmark, which issued writes to an Apache Cassandra database, and the Zstandard compression benchmark using compression level 3, long mode, respectively. This was likely due to the WFQ scheduler's simpler mechanism for rebalacing tasks making less optimal decisions for these benchmarks. Interestingly, we also saw speedup for \system{} on some benchmarks. Overall, the geometric mean of the performance differences between the schedulers across the benchmarks was 0.74\%.

We also ran these benchmarks on ghOSt. The SOL minimal FIFO scheduler described in the ghOSt paper is much slower and does not run some of the benchmarks. A new ghOSt-based weighted fair queueing scheduler is in progress, and runs correctly on all but one benchmark. It is slower on most benchmarks than \system{}. However, since it is not finished, a quantitative comparison would not be fair to ghOSt at this time.

\subsection{Shinjuku Scheduler}
\begin{figure*}[ht]
    \centering
    \begin{subfigure}[b]{0.31\textwidth}
        \includegraphics[scale=0.6,clip]{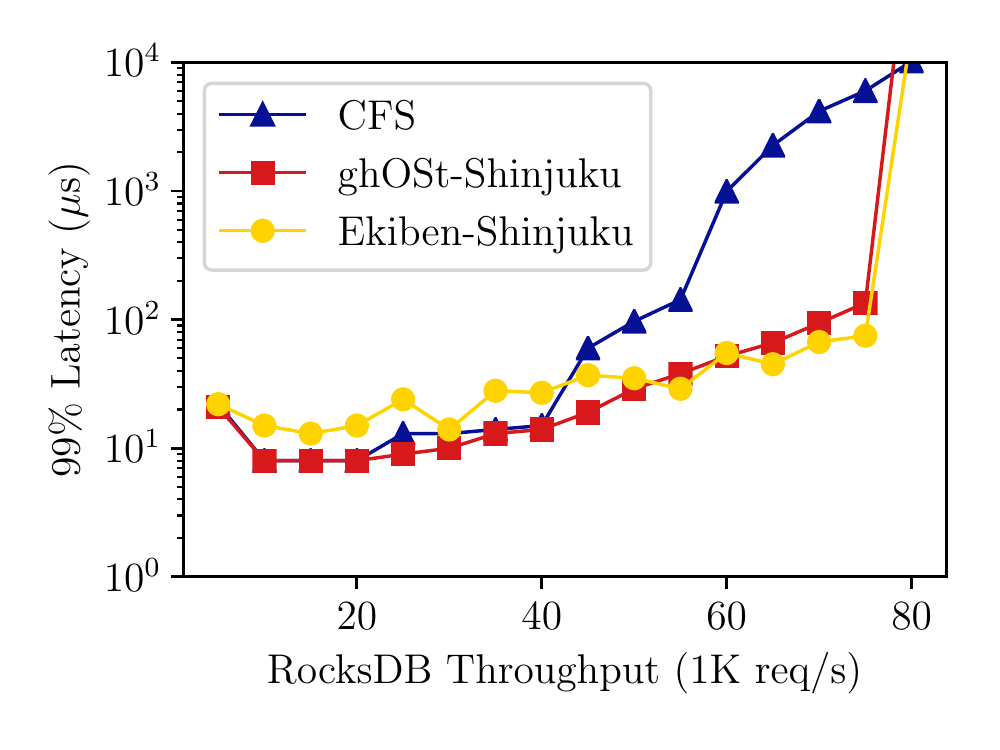}
        \caption{Tail latency for a dispersive load on RocksDB.}
        \label{fig:shinjuku-one-app}
    \end{subfigure}
    \hfill
    \begin{subfigure}[b]{0.31\textwidth}
        \includegraphics[scale=0.6,clip]{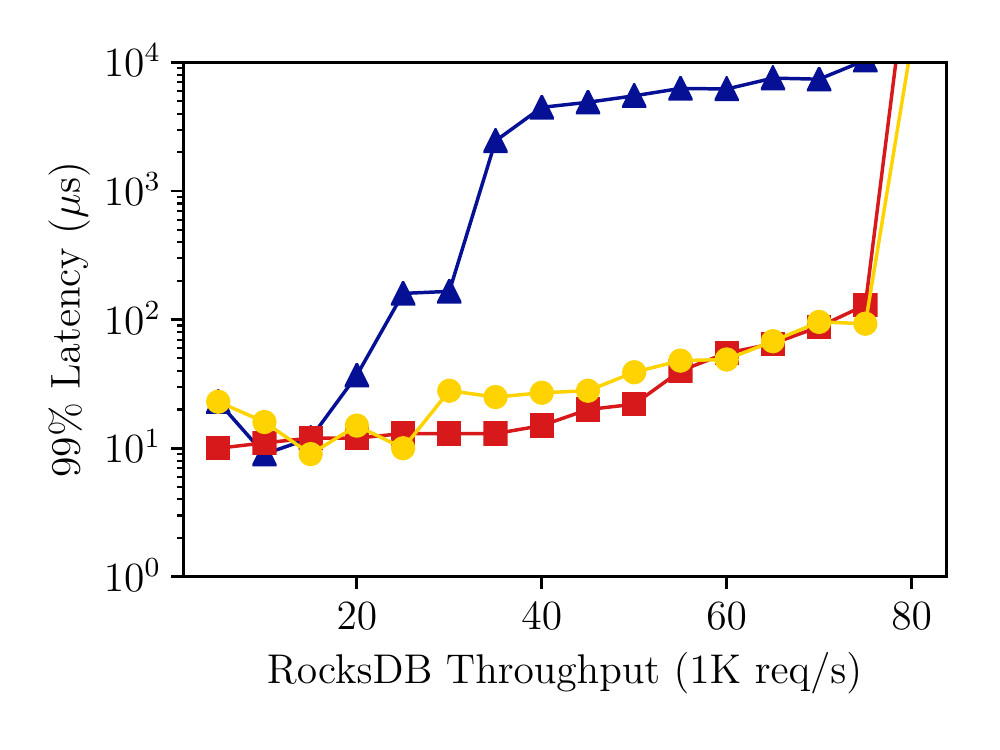}
        \caption{Tail latency for a dispersive load on RocksDB co-located with a batch app.}
        \label{fig:shinjuku-two-app}
    \end{subfigure}
    \hfill
    \begin{subfigure}[b]{0.31\textwidth}
        \includegraphics[scale=0.6,clip]{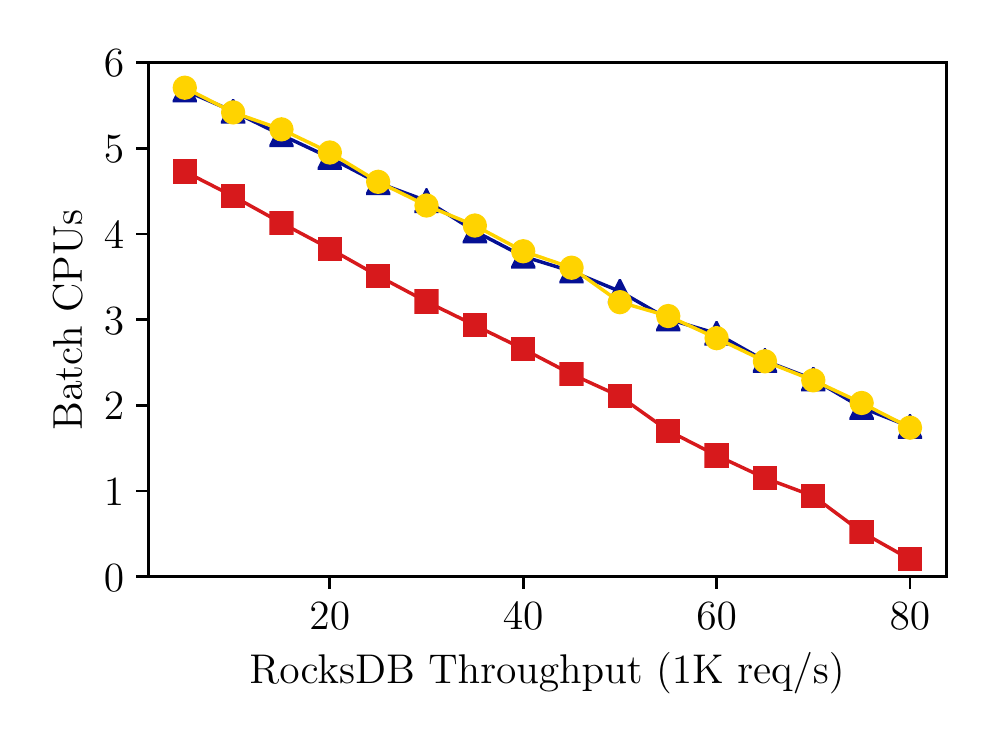}
        \caption{CPU share for a batch application co-located with RocksDB.}
        \label{fig:shinjuku-batch}
    \end{subfigure}
    \caption{Comparison of the \system{} Shinjuku ($\mu$s scale preemption) scheduler compared to CFS and the ghOSt Shinjuku scheduler using RocksDB with and without a batch application. Latency graphs are in log scale. \vspace{-5mm}}
    \label{fig:shinjuku-all}
\end{figure*}
To evaluate our version of the Shinjuku scheduler, we ran the RocksDB benchmarks used in the Shinjuku and ghOSt papers. These benchmarks send queries to an in-memory RocksDB database, with 99.5\% GET requests and 0.5\% range queries. Replicating how this benchmark was run in ghOSt, each GET is assigned to take 4$\mu$s and each range query to take 10ms. If RocksDB responds too quickly, the benchmark spins until the assigned time has elapsed. Three cores were reserved, one for background tasks, one for the load generator, and one for the scheduler if required. The load generator passes tasks to a total of 50 workers running on the other five cores. We evaluate against the ghOSt version of the Shinjuku scheduler. We could not compare against the original Shinjuku scheduler because it requires a specific NIC that we do not have. Both the ghOSt Shinjuku and \system{} Shinjuku schedulers use a preemption timer of 10$\mu$s. The results are shown in \autoref{fig:shinjuku-all}.

In the first benchmark, shown in \autoref{fig:shinjuku-one-app}, only the RocksDB application is run. Note that the y-axis is in log scale. Both he \system{} and ghOSt Shinjuku schedulers achieve low tail latency at high loads due to the short preemption timer; long running range query tasks are preempted quickly, allowing the short GET queries to run. On CFS, tasks run for much longer before being preempted, by default 750$\mu$s, so the GET queries spent more time waiting to be run.

In the second benchmark, a batch application was co-located with RocksDB.
For the CFS and \system{} tests, the batch application was run using CFS. RocksDB is given a priority of -20 while the batch application is 19 (lower is higher priority). 
The batch application is run on ghOSt using a lower priority than RocksDB.
\autoref{fig:shinjuku-two-app} shows the tail latency of RocksDB, and \autoref{fig:shinjuku-batch} shows the CPU share of the batch application.

Both the \system{} and ghOSt Shinjuku schedulers achieve similar tail latency as with no background task because they give priority to the RocksDB workers, with a small time slice. The tail latency on CFS worsens with addition of the batch task.

The batch application receives a similar share of the CPU on CFS and the \system{} Shinjuku scheduler. CFS itself shares cycles between the RocksDB workers and the batch application according to their niceness values. When there are no RocksDB requests the \system{} scheduler seamlessly cedes cycles to CFS to run the batch application. The ghOSt Shinjuku scheduler provides substantially less CPU time to the batch application because it must pay the overhead of the userspace scheduler. In \system{} and CFS, the scheduler is run on the same core as the application as part of regular kernel calls. All of the schedulers give the batch task a much higher CPU share than the original Shinjuku scheduler would~\cite{ghost}.
\begin{table}[t]
    \centering
    \begin{tabular}{l|c|c|c|c}
        \Xhline{4\arrayrulewidth}
        \textbf{Latency} & \textbf{CFS} & \textbf{CFS One Core} & \textbf{Random} & \textbf{Hints}\\
        \hline
        \normalsize
        50th ($\mu$s) & 33 & 17 & 46 & 2\\
        99th ($\mu$s) & 50 & 32032 & 49 & 4 \\
        \Xhline{4\arrayrulewidth}
    \end{tabular}
    \caption{Wakeup latency for the schbench benchmark with two message threads and two worker threads per message thread. Thread wakeup latencies were in microseconds. All runs used a 5s warmup time and ran the benchmark for 30s.\vspace{-3mm}}
    \label{tab:locality_results}
\end{table}

\subsection{Locality Aware Scheduler}

To evaluate the effectiveness of using hints in our locality aware scheduler, we used a modified version of the schbench benchmark. This benchmark starts a specified number of message threads and worker threads. Each message thread and its worker threads send messages back and forth. The benchmark records the wakeup latency of the worker threads to evaluate scheduler overhead. When a message thread and its worker threads are on the same core, wakeup latency can be very low. However, the benchmark uses a futex to wait which does not set the WF\_SYNC flag when waking the workers, so Linux can not detect this pattern~\cite{amd-sched}. In our modified version of the benchmark, we send hints to an \system{} locality aware scheduler to co-locate the message thread with its workers, but place each set of message and worker threads on a different core. We compare this approach to CFS and to the locality aware scheduler with random placement (no hints) as baselines. We also compare to CFS using cgroups to test if the flexibility provided by the hints is necessary for a performance benefit. Cgroups enable specifying a set of cores that a process should run on, but do not support different sets of cores for different threads within the same process. We use cgroups to place all the threads on one core.

The results are shown in \autoref{tab:locality_results}. We use two message threads and two worker threads per message thread.
CFS and the locality aware scheduler with random placement perform similarly because both spread tasks across cores. The locality aware scheduler with hints achieves significantly lower 99\% latency.
Using cgroups to put all threads on one core improves median latency at a cost of much worse tail latency due to the added competition between threads.

\subsection{Arachne Scheduler}
\begin{figure}
    \centering
    \includegraphics[scale=0.69,clip]{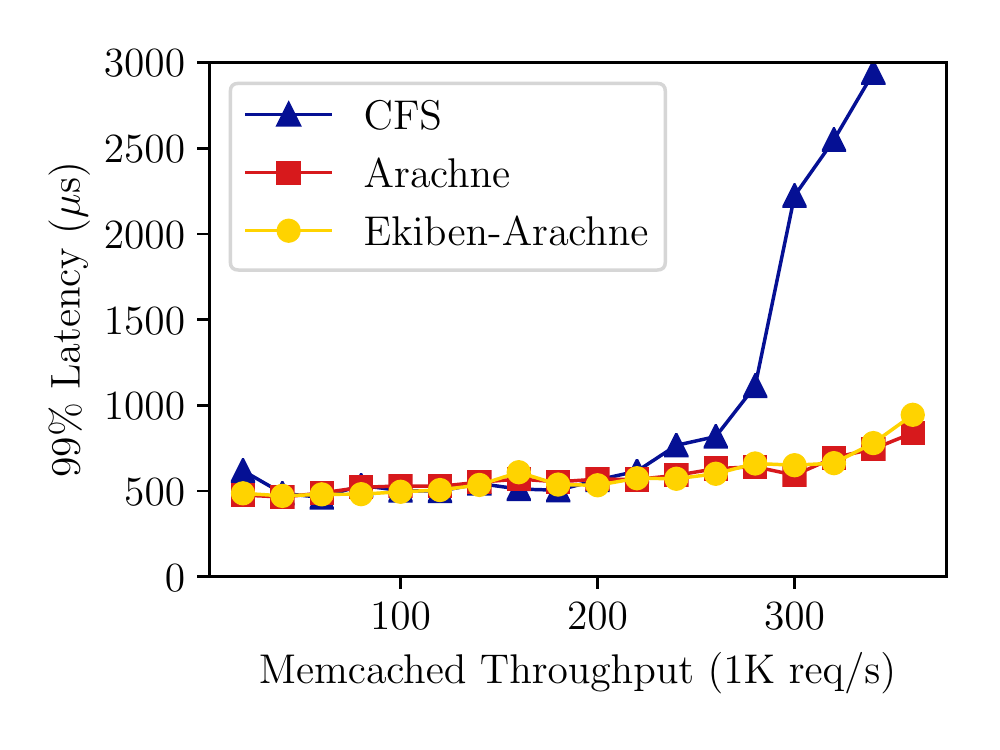}
    \caption{Tail latency of requests to a Memcached server using Mutilate comparing baseline Memcached using CFS to a version using Arachne and a version using Arachne modified to use an \system{} core arbiter.\vspace{-5mm}}
    \label{fig:my_label}
\end{figure}

We evaluate the \system{} version of the Arachne scheduling using a version of the Realistic memcached workload from the Arachne paper~\cite{arachne}. We use the Mutilate benchmark utility~\cite{Mutilate} to generate load for the memcached server, using the key size and distribution, value size and distribution, and inter-arrival distribution of the Facebook ETC workload~\cite{etc}, 1 million records, and 3\% updates. Four clients are used to generate load, enough to make the benchmark server-bound in our tests. Each client creates 16 threads and four connections per thread. An extra client is used to evaluate latency. We compare the \system{} version of Arachne against the original Arachne code and a baseline version of memcached running on CFS. Both of the Arachne versions automatically scale between two and seven cores, reserving one for background tasks. The baseline version uses all eight cores on the machine.

The \system{} version of Arachne achieves similar performance to the original Arachne scheduler, better than CFS at high load.

\subsection{Live Upgrade}
We evaluate the performance impact of live upgrade using the schbench benchmark and timing instrumentation in the kernel. We track the tail latency of schedule operations before, during, and after the upgrade. The upgrade interruption was too short to affect the tail latency of the schbench operations, so we repeated the evaluation with timing calls inserted in the kernel. We evaluate this benchmark on both the one socket machine using 2 message threads and 2 workers per message thread and the two socket machine with 2 message threads and both 2 and 40 workers per message thread. On the one socket machine, the upgrade takes 1.5$\mu$s. On the two socket machine, the upgrade takes 9.9$\mu$s and 10.1$\mu$s for 2 and 40 workers, respectively.  All were averaged over three runs.


\subsection{Record and Replay}

We evaluate the performance of record and replay using the \texttt{perf bench sched pipe} benchmark on the WFQ scheduler. This benchmark completes in around 4 seconds during regular operation. During record, the benchmark completes in around 30 seconds, and the replay takes around 3 minutes. Record is slower than regular operation because all relevant operations must be sent to the record program, which writes a file asynchronously on a different core.

During replay, the first 30 seconds are spent reading the file and parsing lock operations. The rest is due to the mechanism for ensuring concurrent operations occur in the same order as the live execution. This benchmark has many calls to the scheduler. In the kernel, these operations are very fast. During replay, if threads arrives in a different order than what was recorded, we need to block the thread and wake it up later to try again.
These constant sleeps and wakes add latency and account for most of the execution time during replay.

\section{Discussion}\label{discussion}

In this section, we report on our experience using \system{} to develop kernel schedulers.
\system{}'s design made scheduler development much more straightforward than it would be in Linux. In fact, two of the schedulers (the Shinjuku and Locality Aware schedulers) were written by undergraduates with no prior experience in Linux kernel development. They were able to start coding with relatively little ramp up and were able to build, test, and debug their code on their own. They primarily benefited from modularity, because they were using \system{} when record and replay and live upgrade had not been implemented.

Being able to use the \texttt{Schedulable} data structure as proof that a task is schedulable on a core allowed us to ensure that we were picking runnable tasks. This helped us identify and address bugs faster because we would see a compile time error from \texttt{pick\_next\_task} returning an incorrect \texttt{Schedulable} thread rather than putting the kernel into a bad state by trying to run a task on the wrong core.

We ran into relatively few runtime bugs, and those bugs were fairly easy to address. Most have been deadlocks, which are painful to encounter because they force a reboot, but proved surprisingly easy to debug, in line with prior OS development experience~\cite{birrell1989an}. Often, relatively few lock operations were touched in between two runs of the code, so finding the relevant changes to lock operations was easy. With more complicated deadlocks, it was easy to look through the scheduler code and check the order of lock acquires because the largest scheduler (WFQ) was only around 600 lines of code.
Of the non-deadlock bugs, most were conceptual, such as a benchmark having tasks yield when we had not yet implemented \texttt{task\_yield} or miscalculating the fair time slice to assign tasks in WFQ.
While fixing these bugs sometimes required rebooting the machine, we never had to recompile the kernel to fix them, so our iteration times were quite short. 




The general design of an interposable, message-passing based interface that uses memory sharing under the hood had a number of benefits. We gained many of the benefits of modularity for development velocity because of clean interfaces and isolation from the rest of the kernel, allowing the developer to focus their energy on the algorithm being implemented.
In its implementation, \system{} uses function calls and memory sharing, limiting the overhead that can come with modular designs. Because \system{} schedulers run in the kernel, the core scheduling code can quickly and easily make synchronous calls to the scheduler, allowing it to quickly respond to changes in state.



\system{}'s design also made it easy to implement additional features. Because all the functionality was contained in the module and \system{}-C contacts the scheduler through a single function pointer, live upgrade is as simple as quiescing the state and replacing the function pointer. Due to Rust's support for generic data types and traits, custom hint data structures could be defined as type parameters on the scheduler and any requirements on the data types can be expressed as trait bounds. \system{}'s design supported record and replay debugging very smoothly. The interposable, message passing based interface and made it simple to record relevant state that was passed into the scheduler or returned from calls into the kernel. Recording nondeterministic behavior is one of the main challenges for record and replay on parallel systems, but using safe Rust made this much simpler. Due to Rust's safety guarantees, we knew that the scheduler could not contain race conditions or any other undefined behavior, so the only sources of nondeterminism were timing and the order of lock acquisitions. All timing state was handled by the kernel and passed into the scheduler, and so was automatically handled by recording the messages. To correctly handle concurrency, we only needed to record and replay the order in which locks were acquired.

\section{Related Work}

\textbf{Scheduler Frameworks. }
GhOSt~\cite{ghost} is the closest analog to \system{}. It allows the user to replace the kernel scheduler with a userspace scheduler.
Calls to the kernel scheduler are forwarded to the userspace scheduler agent, which responds with decisions that are then applied in the kernel. The kernel does not wait for decisions from the userspace scheduler to schedule a task, instead applying decisions asynchronously at a later call into the kernel scheduler.
GhOSt provides good development velocity but can add significant overhead and scheduling latency, and the asynchronous model makes it difficult to precisely mirror kernel scheduler decisions.
Mvondo et.al.~\cite{kernel-sched} also proposed a framework for implementing new schedulers in Linux, inspired by how eBPF enables developers to implement new networking functionality for the kernel. As far as we can tell, this system has not been built yet.
AMD engineers have proposed a method for providing userspace hints to a kernel scheduler~\cite{amd-sched}, but only a small number of predefined types of hints are supported.



\textbf{Safe Languages in OS Development. }
It is common for research operating systems to be written entirely in a type-safe high-level language. Examples include Pilot~\cite{pilot}, SPIN~\cite{spin}, Singularity~\cite{singularity}, Biscuit~\cite{biscuit}, Redox~\cite{redox}, Tock~\cite{tock}, Theseus~\cite{theseus}, and RedLeaf~\cite{redleaf}.
None of these target improving development velocity for a widely used commercial operating system.
These approaches show that benefits can be gained by using safe languages, but all involve rewriting the entire operating system.
We are also not the first to integrate Rust into the Linux kernel~\cite{rust_devices,linux-module-rust,bento}, but these projects do not provide safe interfaces for schedulers. Most similar to our work is Bento~\cite{bento}. Bento takes a similar approach to \system{} but applied to file systems. Because it targets file systems and sits behind the kernel page cache, Bento can tolerate much higher latency than \system{}. 

The Berkeley Packet Filter (eBPF)~\cite{bpf} enables safe extensibility in Linux using a restricted, type safe language and programming environment.
eBPF programs can only be inserted at specific points in the kernel and are verified by the kernel before being run.
GhOSt~\cite{ghost} extends eBPF to enable implementing part of the scheduler using eBPF. eBPF programs can be updated at runtime, but the locations where code is inserted cannot be.
To ensure safety of user code running in the kernel, eBPF is quite restrictive. Programs are limited to only running at pre-defined points in the kernel, only calling whitelisted functions, and not having unbounded loops. For example, it would be difficult to express a red-black tree or B-tree in eBPF, as used in CFS and WFQ.

\textbf{Live Upgrade. }
Live upgrade in \system{} is most similar to the techniques used in Bento~\cite{bento}, but applied to schedulers.
Plugsched~\cite{plugsched} does live upgrade of the entire scheduler subsystem of the Linux kernel without modifying any kernel code using code analysis to detect boundaries of the scheduler subsystem and stack analysis to maintain consistent state across the upgrade.
Other tools for live upgrade of Linux systems include ksplice~\cite{ksplice-paper,ksplice}, kpatch~\cite{kpatch}, or kGraft~\cite{kgraft}.
These replace individual functions with new implementations, and are primarily used for security patches that do not modify kernel data structures.
Other research systems, such as K42~\cite{k42}, PROTEOS~\cite{proteos}, LUCOS~\cite{lucos}, and DynAMOS~\cite{dynamos}, support upgrading complex, modular components in either new operating systems (K42 and PROTEOS) or in Linux using shadow data structures or virtualization (LUCOS and DynAMOS).
Our work enables live upgrade of large modules in a commodity operating system with state transfer by introducing a framework layer that handles quiescing state.


\textbf{Running in Userspace. }
Running operating system services in userspace can also increase development velocity of operating system components~\cite{mach,l4}, though this can sacrifice performance.
Most similar to \system{} is GhOSt~\cite{ghost}.
Another approach is to run the component in userspace on dedicated cores with communication through shared memory queues~\cite{urpc,barrelfish,tas,snap}. 
This approach has been used for implementing new network stacks~\cite{tas,snap} and scheduler algorithms~\cite{shinjuku,arachne}.
With direct access to hardware devices, performance can often be competitive with an equivalent kernel implementation.
However, interaction with the rest of the kernel can be restricted using this approach, possibly limiting the ability for any part of the system to make global decisions~\cite{interposition}.


\textbf{Record and Replay. } Record and replay systems enable debugging code by first recording a trace of operations and then later replaying those operations against the target code, enabling debugging of the recorded trace. These systems require instrumentation around the target code in order to record the necessary data.
There are a variety of approaches to record and replay, from recording the whole system~\cite{revirt,retrace} to using kernel instrumentation to record user programs~\cite{recplay,eidetic,castor} to running both the target code and the instrumentation in userspace~\cite{jockey,liblog,pinplay,idna} to instrumenting language and application runtimes~\cite{dejavu,timelapse}. One project~\cite{concolic-execution} specifically records and replays kernel modules using mechanisms for whole binary analysis and instrumentation of kernel module interfaces.
As far as we know, we are the first project to selectively record a kernel module, as opposed to the whole kernel, and to replay the behavior at userspace on the same code as ran in the kernel.
There is a long history of work on deterministic record and replay on multicore systems to ensure that data accesses are performed in the same order during record and replay~\cite{odr,recplay,r2,retrospect,bugnet,instantreplay,leap}. We combine content-based message passing recording, where the content of all messages across a message passing interface are recorded, with a software only shared memory recording scheme, where synchronization and access to shared state are recorded. Unlike other shared memory recording schemes, we do not have to detect and record race conditions due to Rust's safety guarantees, so we record only synchronization accesses.

\section{Conclusion}
This paper presents \system{}, a framework for rapid development of high performance Linux kernel schedulers.
\system{} enables safe, high performance kernel schedulers with seamless live upgrade, bidirectional user communication channels, and record and replay debugging.
A scheduler implemented with \system{} is able to achieve similar performance to CFS, the default Linux scheduler, on a wide range of benchmarks.
Other \system{} schedulers mimic recent research schedulers, but integrated with Linux.
\system{}'s schedulers can be upgraded with only 10.1$u$s of service interruption, and the record and replay debugging allows for slow but functional userspace debugging of the kernel scheduler code.

\clearpage

\bibliographystyle{ACM-Reference-Format}
\bibliography{main}
\clearpage
\appendix
\section{Appendix}

\begin{table}[h!]
    \centering
    \begin{tabular}{l|l}
        \Xhline{4\arrayrulewidth}
        Arrayfire, 1 & Arrayfire, BLAS CPU \\
        Arrayfire, 2 & Arrayfire, Conjugate Gradient CPU \\
        Cassandra, 1 & Cassandra, Writes \\
        ASKAP, 4 & ASKAP, Hogbom Clean OpenMP \\
        Cpuminer, 2 & Cpuminer, Triple SH-256 Onecoin \\
        Cpuminer, 3 & Cpuminer, Quad SHA-256 Pyrite \\
        Cpuminer, 4 & Cpuminer, Myriad-Groestl \\
        Cpuminer, 6 & Cpuminer, Blake-2 S \\
        Cpuminer, 11 & Cpuminer, Skeincoin \\
        Ffmpeg, 1, 1 & Ffmpeg, libx264, Live \\
        Graphics-Magick, 4 & Graphics-Magick, Resizing \\
        OIDN, 1 & OIDN, RT.hdr\_alb\_nrm.3840x2160 \\
        OIDN, 2 & OIDN, RT.ldr\_alb\_nrm.3840x2160 \\
        OIDN, 3 & OIDN, RTLightmap.hdr.4096x4096 \\
        Rodina, 3 & Rodina, OpenMP Leukocyte \\
        Zstd, 2 & Zstd, 3 Long Mode Compression \\
        Zstd, 4 & Zstd, 8 Long Mode Compression \\
        AVIFEnc, 4 & AVIFEnc, 6 Lossless \\
        Libgav1, 1 & Libgav1, Summer Nature 1080p\\
        Libgav1, 2 & Libgav1, Summer Nature 4k \\
        Libgav1, 3 & Libgav1, Chimera 1080p \\
        Libgav1, 4 & Libgav1, Chimera 1080p 10-bit \\
        OneDNN, 4, 1 & OneDNN, IP Shapes 1D, f32 \\
        OneDNN, 5, 1 & OneDNN, IP Shapes 3D, f32 \\
        OneDNN, 7, 1 & OneDNN, RNN Training, f32 \\
        OneDNN, 7, 2 & OneDNN, RNN Training, u8s8f32 \\
        OneDNN, 7, 3 & OneDNN, RNN Training, bf16bf16bf16 \\
        \Xhline{4\arrayrulewidth}
    \end{tabular}
    \caption{Full names of the Phoronix Multicore benchmarks presented in the paper.}
    \label{tab:app_description}
\end{table}

\subsection{WFQ Functional Equivalence}
We run several benchmarks to evaluate whether our \system{} WFQ scheduler correctly implements the expected behavior of a WFQ scheduler. We compare the \system{} WFQ scheduler to CFS, also a WFQ scheduler, to ensure that our scheduler implements equivalent behavior.

One benchmark evaluates whether the scheduler fairly shares CPU time among ready tasks. This benchmark started five tasks that each perform CPU-intensive work, specifically repeatedly adding to a value. By default these tasks are placed on different cores. We expect these tasks to complete in roughly the same amount of time due to the lack of competition. We then force the schedulers to place all of the tasks on the same core.
We saw the expected behavior. On both schedulers, the tasks took around 4.6 seconds to complete when the tasks are not co-located. When the tasks were co-located, the tasks took around 22.2 seconds.

We also evaluate how the schedulers handle weighting by reducing one of the five tasks to minimum priority. We would expect the lowest priority task to complete slower than the others; the other tasks should share time fairly and complete in the same amount of time. Again, we saw the expected behavior on both schedulers. When one task was reduced to the minimum priority, all four other tasks complete in around 17.6 seconds and the lowest priority task completes in 4.4 seconds later.

A third benchmark tests task placements. We start one task per core with each task performing CPU-intensive work, specifically repeatedly adding to a value. We expect both schedulers to default to placing each task on a separate core. We then force one task to change cores. With no movement, each task completes in around 9 seconds on both schedulers and there is very low variation in task runtime. When a task is moved, all tasks still complete in around 9 seconds on both schedulers. CFS shows roughly the same variation in task runtime even when a task was moved. The \system{} WFQ scheduler shows higher standard deviation in task runtimes when a task is moved, from 0.001s to 0.018s, because it takes longer to move a task due to its less sophisticated methods of rebalancing tasks.


\end{document}